\documentstyle[pra,aps,amsfonts,epsfig]{revtex}
\begin{document}
\draft
\twocolumn[\hsize\textwidth\columnwidth\hsize\csname 
@twocolumnfalse\endcsname
\title{Measurement of the degree of Polarization Entanglement Through Position Interference} 
\author{M. Fran\c ca Santos$^{1*}$, P. Milman$^{1,2}$, A. Z. Khoury$^{3}$ and P.H. Souto Ribeiro$^{1}$}
\address{$^{1}$Instituto de F\'{\i}sica, 
Universidade Federal do Rio de 
Janeiro, Caixa Postal 68528, Rio de Janeiro, RJ 21945-970, Brazil\\
$^{2}$ Laboratoire Kastler Brossel, D\'epartement de Physique de L'Ecole Normale Sup\'erieure, \\24 rue Lhomond, Paris Cedex 05, France\\
$^{3}$Instituto de F\'{\i}sica, 
Universidade Federal Fluminense, BR-24210-340 Niteroi, RJ, Brazil} 
\date{\today}
\maketitle
\begin{abstract}
We produce polarization entangled states with variable degree of entanglement for twin photons. Entanglement in polarization is coupled to entanglement in position that produces transverse coincidence interference fringes. We show both theoretically and experimentally that, due to this coupling, 
we can use the interference pattern to measure the polarization degree of entanglement.
\end{abstract}
]

\section{Introduction}

Entanglement is the central property behind the experiments performed with twin photons produced in the parametric down-conversion process\cite{1}. A large number of exciting applications of 
fundamental properties of light such as quantum nonlocality, wave-particle duality, quantum interference and many others, all of them closely related to entanglement, have already been performed in such systems. More recently, entangled photon pairs have also proven to be an adequate environment to study quantum information theory. The non-local correlations between the signal and idler photons allow 
the implementation of fundamental quantum mechanic experiments, such as teleportation~\cite{2} and 
quantum erasers~\cite{3}, as well as more practical applications like quantum cryptographic 
devices~\cite{4}.

Entangled states of twin photons involving the correlation between many different degrees 
of freedom can be easily generated and detected. Particularly, polarization entanglement has been 
extensively utilized, not only because it is easily produced and controlled~\cite{5,6}, but also because 
it is defined in a two-dimensional Hilbert space similar to spin-1/2 systems. These
systems are suitable for implementing quantum computation and communication~\cite{7} schemes, and EPR experiments~\cite{5,8}. However, polarization is not the only degree of freedom that can present entanglement in this system. Energy and momentum~\cite{9}, for instance, can also present non-local correlations to be explored. 

In this paper, we present an experiment and a theoretical approach describing the coupling between two 
different degrees of freedom of the detected pairs: polarization and transverse momentum. Two sources of photon pairs, i.e., two nonlinear crystals pumped by the same laser beam, are placed collinearly in two 
different positions in space. Many quantum numbers are associated with the twin-photons emitted in each 
one of the crystals but we will be interested, particularly, in their polarization and their position of 
creation. Each crystal generates pairs with defined polarization depending on their relative orientation 
with respect to the pump laser polarization. Furthermore, the twin-photon beams have distinct transverse 
components of momentum. However, the proper alignment of these beams combines them to make the 
origin of creation of each detected pair indistinguishable to position measurements. Therefore, we will 
call it position entanglement~\cite{10}, as we shall see bellow.

The position entanglement generated by such system leads to quantum interference that can be observed 
through coincidence patterns~\cite{10}. Since the origin of each pair and their polarization are 
intrinsically correlated we have utilized position interference for monitoring the degree of polarization 
entanglement. Even though this degree of entanglement can be measured by other means, like quantum 
state tomography\cite{6}, the experiment we present stresses the coupling between entanglement in two 
different degrees of freedom. It will also be shown that, as in a 
quantum eraser experiment, the coupling between these systems can be controlled, leading to the 
increase or decrease of fringe visibility.

\section{Polarization and position entanglement}

\begin{figure}[h]
\vspace*{5cm}
\epsfig{file=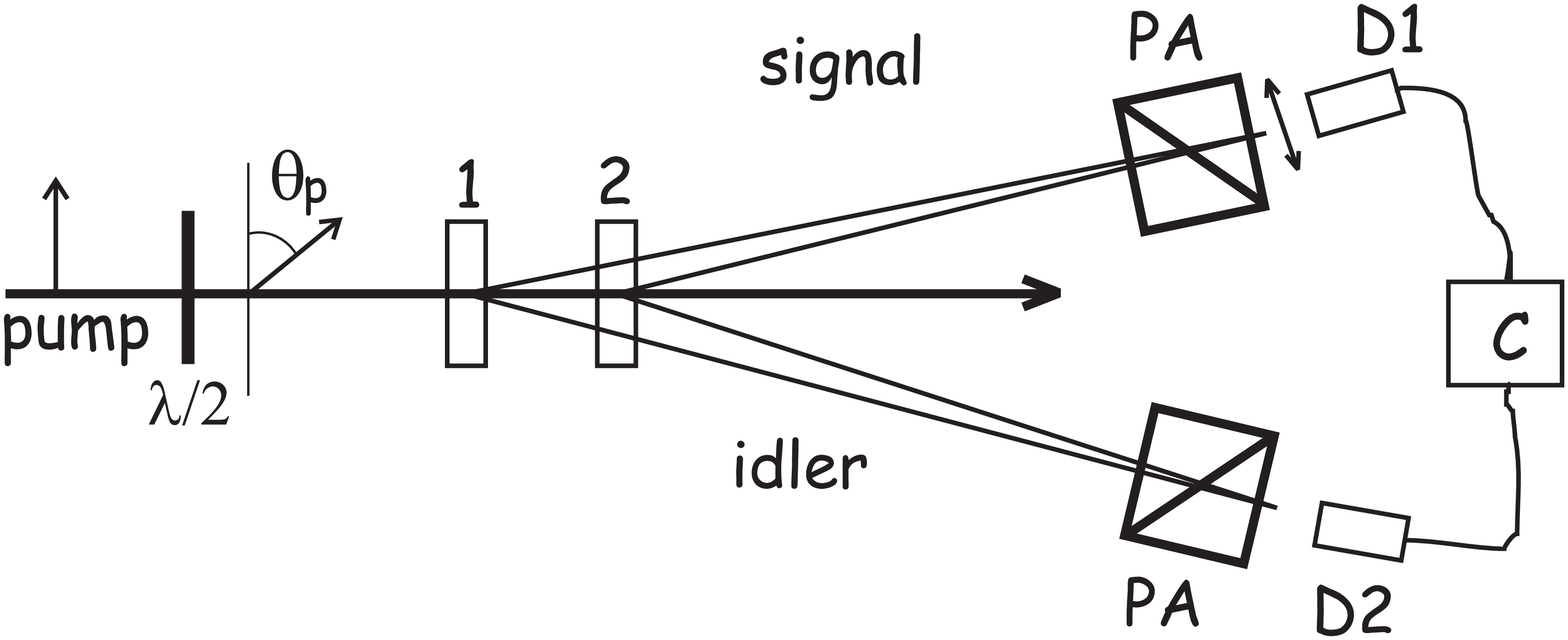,width=8cm,height=5cm}
\caption{Outline of the experiment. 1 and 2 are nonlinear crystals, PA is polarization
analyzer, D1 and D2 are signal and idler detectors respectively.}
\label{fig1}
\end{figure}

A sketch of the experimental set-up is shown in Fig.\ref{fig1}. Two nonlinear crystals produce twin 
photon pairs. All photons have the same wavelength
but their polarizations depend on the orientation of the optical axes of the crystals relatively to the 
pump beam polarization direction.

When both crystals produce photon pairs with the same polarization, we have quantum interference 
between two different but indistinguishable origins of the two-photon wave field; either crystal 1 or 
crystal 2. For this superposition to be effective signal and idler beams generated in each crystal must 
be adequately aligned, i.e., there must be good spatial mode matching between them.   
Due to the long coherence length of the pump laser, the origin of the photons arriving at the coincidence
detectors, whether it is the first or the second crystal, cannot be determined in principle. In this case, 
the system presents position entanglement. 
It is, then, possible to measure position interference in a double-slit type experiment. 
Each crystal plays the role of one slit, emitting a photon pair instead of a single photon as in 
usual Young's fringes experiments. If we move one (or both) of the detectors in the plane 
perpendicular to signal or idler beams we get Young's fringes in the coincidence counting rate. The 
details of this particular experiment are described elsewhere~\cite{10}.

For the purpose of this paper, it is enough to consider a monomode theory.
In this case, the most general state of the photons arriving at the detectors is 
\begin{equation}
\left| \Psi _{\phi_1,\phi_2}\right\rangle =\alpha \left| 2_{\phi_1},0_{\phi_2}\right\rangle
+e^{i\varphi } \beta \left| 0_{\phi_1},2_{\phi_2}\right\rangle ,  
\label{est1} 
\end{equation}
where $\phi_1,\phi_2$ indicates the polarization of the photons produced in crystal $1$ and crystal $2$ respectively. $ \varphi $ is a phase which depends on the optical path difference from crystal $1$ and crystal $2$ to the detectors,  
\begin{equation}
\varphi = k(\Delta x_1-\Delta x_2)+\varphi_0,
\label{est1a}
\end{equation}
where $\Delta x_{1,(2)}$ is the distance from crystal 1 (2) to the detector 1 plus the distance from crystal 1 (2) to the detector 2. k is the wavenumber of all fields and $\varphi_0$ is a constant phase. $\varphi$ also depends on some other factors (that do not vary in our experiment) as the phase accumulated by the pump
between crystals, all contained in the constant phase $ \varphi_0 $. At the peak of the interference pattern $\varphi =0$. $|\alpha|^2$ ($|\beta|^2$) gives the probability of finding a pair created in crystal $1$ ($2$) and it depends on the projection of the pump beam polarization onto its optical axis direction.

This is a rich system presenting interchangeable polarization and position entanglements. Depending on the orientation of the crystals and the polarization of the pump laser many different states can be derived from (\ref{est1}). For example, if $\phi_1 = \phi_2 = \phi$, both crystals produce pairs with
the same polarization and with the same 
probability so that state (\ref{est1}) is reduced to the usual Young's experiment \cite{11a} state 
\begin{equation}
\left| \Psi _{\phi,\phi}\right\rangle =\frac{\left| 2,0\right\rangle+e^{i\varphi }\left| 0,2\right\rangle}{\sqrt{2}} ,  
\label{est2} 
\end{equation}
We dropped the polarization index because, in this case, this degree of freedom factorizes. It is impossible to determine from which crystal each detected pair came and state (\ref{est2}) can be viewed as a Bell state for position. In ideal conditions, i.e., perfect mode matching, this state presents interference leading to coincidence interference patterns with visibility equal to one. Visibility is defined by
\begin{equation}
\mu=\frac{C_M-C_m}{C_M+C_m}, 
\label{est2a}
\end{equation}
where $C_M$ is the maximum and $C_m$ is the minimum of the coincidence interference curve. 

Another interesting situation is found when the crystals produce pairs at orthogonal polarizations $\phi_1 = H$ and $\phi_2 = V$ and the pump laser is oriented midway from them, 
$\theta_p = (H + V)/\sqrt{2}$(or $\theta_p = 45^{0}$). 
In this case, the photons are tagged with an origin identifier and Young's interference fringes have zero visibility. On the other hand, it is now impossible to determine the polarization of the detected photon pair and state (\ref{est1}) reduces to the familiar polarization Bell state:
\begin{equation}
\left| \Psi _{H,V}\right\rangle =\frac{\left| HH\right\rangle+e^{i\varphi }\left| VV\right\rangle}{\sqrt{2}}.  
\label{est3} 
\end{equation}
Different orientations of the pump laser lead to non-maximally entangled states \cite{6} described by
\begin{equation}
\left| \Psi _{H,V}\right\rangle =\alpha\left| HH\right\rangle+e^{i\varphi }\beta\left| VV\right\rangle.  \label{est4} 
\end{equation}

Most generally, this system presents position and polarization entanglement, 
as it is described by state (\ref{est1}). The visibility of the interference fringes is given by 
$\mu = 2|\alpha||\beta|cos(\phi_2-\phi_1)$ and depends on both the polarizations and the probability to generate pairs in each crystal. Notice that, for state (\ref{est2}), $\phi_1 = \phi_2$ and $\alpha = \beta$, so that $\mu = 1$. This state is maximally entangled in position and it has no polarization entanglement. On the other hand, for state (\ref{est3}), $\phi_1-\phi_2 = \frac{\pi}{2}$ and $\mu = 0$, i.e. when $\alpha = \beta$ we have maximal polarization and no position entanglement. Rotation of the relative orientation of the crystals changes the entanglement of the detected pairs from position to polarization, which can be observed by measuring the position interference. 

Another controlling variable for the polarization degree of entanglement is the orientation of the pump beam polarization ($\theta_p$). Its influence can be observed by placing a polarization analyzer in front of each detector. If we set the crystals at orthogonal polarizations and orient the analyzers at $45^0$ with respect to them, we regain a position interference because it is no longer possible to determine the origin of each detected pair, although we know their polarization for sure.  In this case we produce position entangled states with variable degree of entanglement,
\begin{equation}
\left| \Psi\right\rangle = \alpha\left|2,0\right\rangle+e^{i\varphi }\beta\left|0,2\right\rangle,  \label{est5} 
\end{equation}
much in the same way as their equivalent in polarization described by state (\ref{est4}). Therefore
this setup is able to change polarization entanglement into position's. The visibility is, then, given solely by $\mu = 2|\alpha||\beta|$ and it will be zero when the polarization of the pump is equal to the orientation of one of the crystals, i.e. either $\alpha$ or $\beta$ will be equal to zero, in which case only one crystal is pumped and we know the origin of each pair for sure. Visibility will be one when both crystals generate pairs with the same probability, $\theta_p = (H + V)/\sqrt{2}$, $\alpha = \beta = 1/\sqrt{2}$.

\section{Experimental results}

A c.w. He-Cd laser operating at 442 nm is utilized to pump two Lithium Iodate crystals as it is shown in Fig.\ref{fig1}. The crystals are 1cm long and cut for collinear degenerate type I down-conversion. In our experimental scheme, they are tilted so that converted beams with the same wavelength emerge from the crystals at about 3 degrees from the pump beam direction. Signal beams produced in crystal 1 and crystal 2 are directed to  detector 1 and idler beams from crystal 1 and crystal 2 are directed to detector 2.
Crystal 2 is placed about 1cm  from crystal 1. The pairs of photons are detected with avalanche photodiode counting modules, which are placed about 1m from crystal 2. The detection scheme also includes a thin slit of about 0.5 mm width, a 10nm bandwidth interference filter centered at 884nm and a 25mm focal length lens before each one of the detectors. Signal and idler beams pass through polarizing beam splitters with their axis oriented at $45^0$ from vertical and  horizontal polarizations, before reaching the detector entrance slit. Detectors are mounted on translation stages that allows scanning the transverse plane of the incoming beams. 

\begin{figure}[h]
\vspace*{5cm}
\epsfig{file=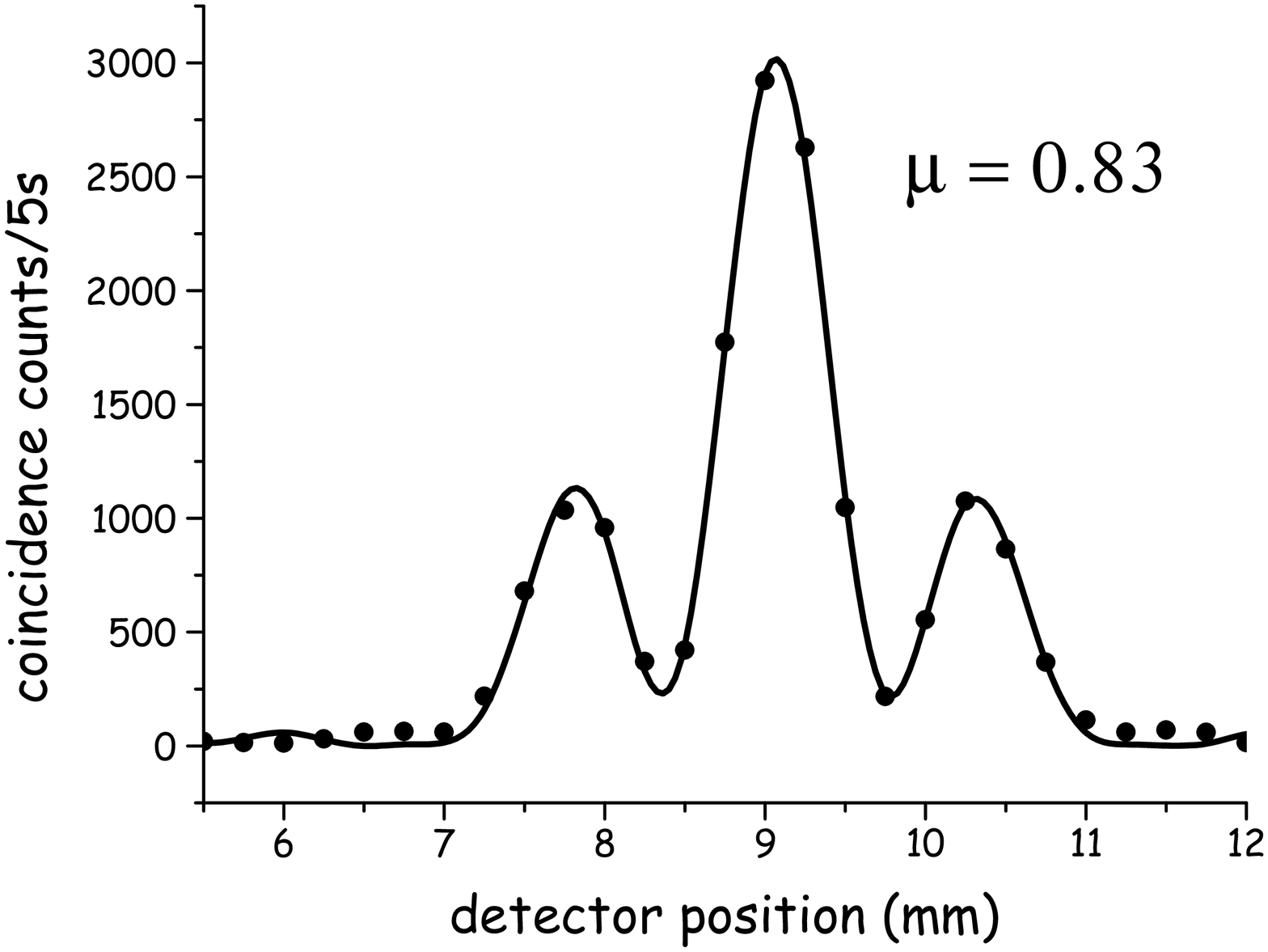,width=9cm,height=5cm}
\caption{Quantum interference for twin beams with the same polarization. Dots are experimental
data. The error bars are negligible. The solid line is a fitting to the usual double slit pattern function
giving the visibility.}
\label{fig2}
\end{figure}

In a first set of measurements, the optical axes of both crystals were vertically oriented. The tilt angle for each crystal is adjusted for obtaining simultaneous coincidence detection counts due to twin photons originated in both crystals for the same position of the signal(det1) and
idler(det2) detectors. In this case, as both crystals generate down-converted beams with the same polarization, their origin is indistinguishable and they present position entanglement, as already explained. It is worth noting that the individual signal or idler photons originated at different crystals are distinguishable, but the detection of the signal(idler) on one side through a small aperture makes the idler(signal) on the other side indistinguishable \cite{11}. As a result we have quantum interference. This leads to interference fringes that can be measured by displacing one (or both) detector in the horizontal direction. In Fig.\ref{fig2} we present a typical interference pattern obtained with position entangled states directly generated by crystals producing twin photons with the same polarization. The fringes were obtained by the displacement of the signal beam detector across the horizontal direction. We would have obtained a similar pattern if we had displaced the idler detector. 
A pattern with doubled frequency would be obtained displacing signal and idler detectors simultaneously.

In a second set of measurements, we turn one of the crystals around the pump beam direction of propagation in order to obtain the state described in Eq. (\ref{est4}). This is the configuration for obtaining polarization entangled photons with variable degree of entanglement introduced by Kwiat et al.\cite{6}. Now, signal and idler photons pass through a 45$^{0}$ polarization analyzer before detection. The pair of photons generated in crystal 1 can be distinguished from the pair generated in crystal 2
because one of them has vertical polarization and the other has horizontal polarization. However, as they cross the 45$^{0}$ polarization analyzers, they become indistinguishable again. Therefore, if a maximally entangled photon pair is produced when the pump beam is polarized at 45$^{0}$, interference fringes in the transverse detection plane must also be detected with a visibility nearly equal to one. On the other hand, if disentangled photon pairs are produced, they will not lead to coincidence interference
fringes in the transverse detection plane.

\begin{figure}[h]
\vspace*{5cm}
\epsfig{file=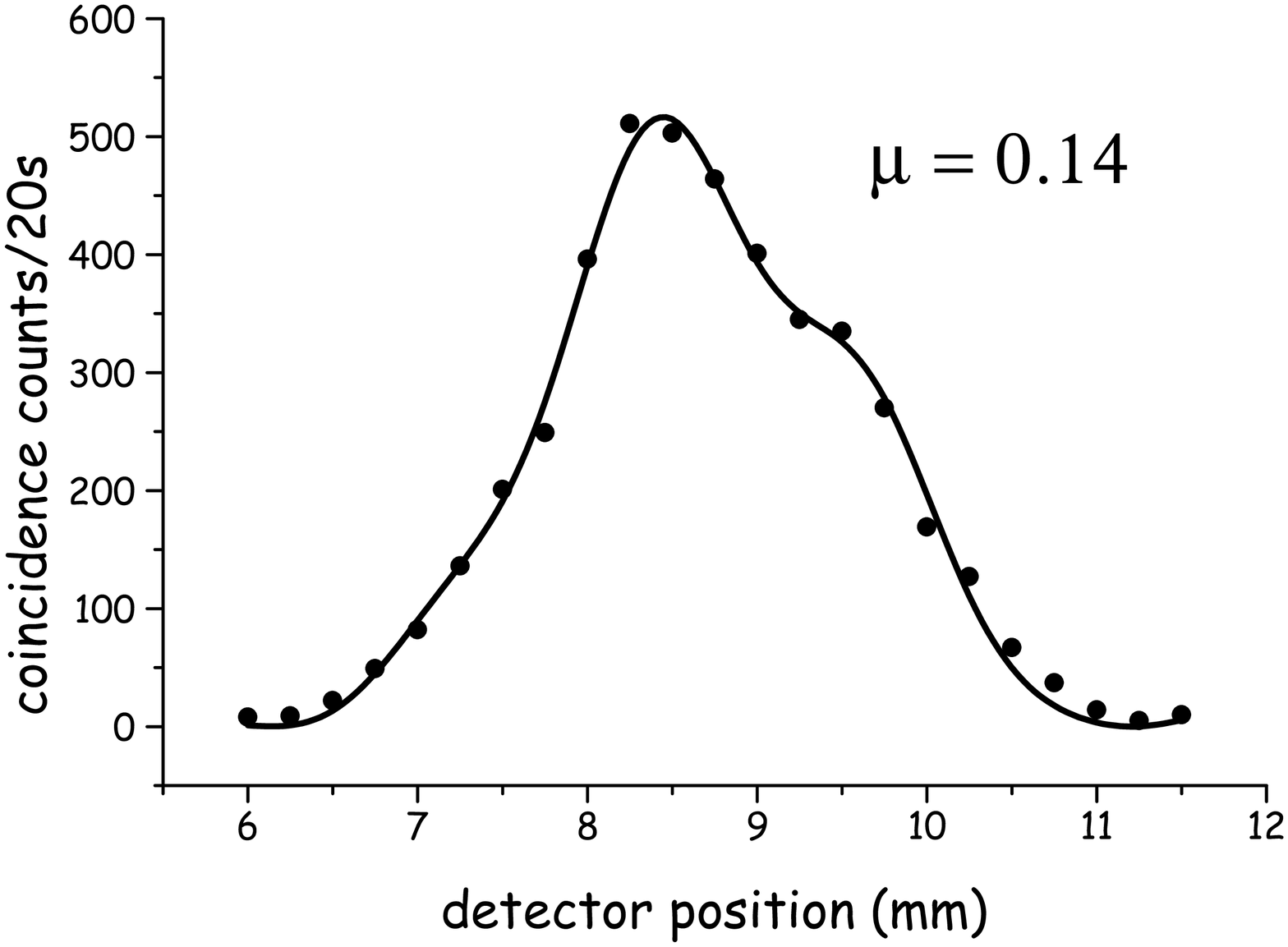,width=9cm,height=5cm}
\caption{Quantum interference for twin beams with different polarizations. Low visibility
and low degree of polarization entanglement. Dots are experimental
data. The error bars are negligible. The solid line is a fitting to the usual double slit pattern function
giving the visibility.}
\label{fig3}
\end{figure}

\begin{figure}[h]
\vspace*{5cm}
\epsfig{file=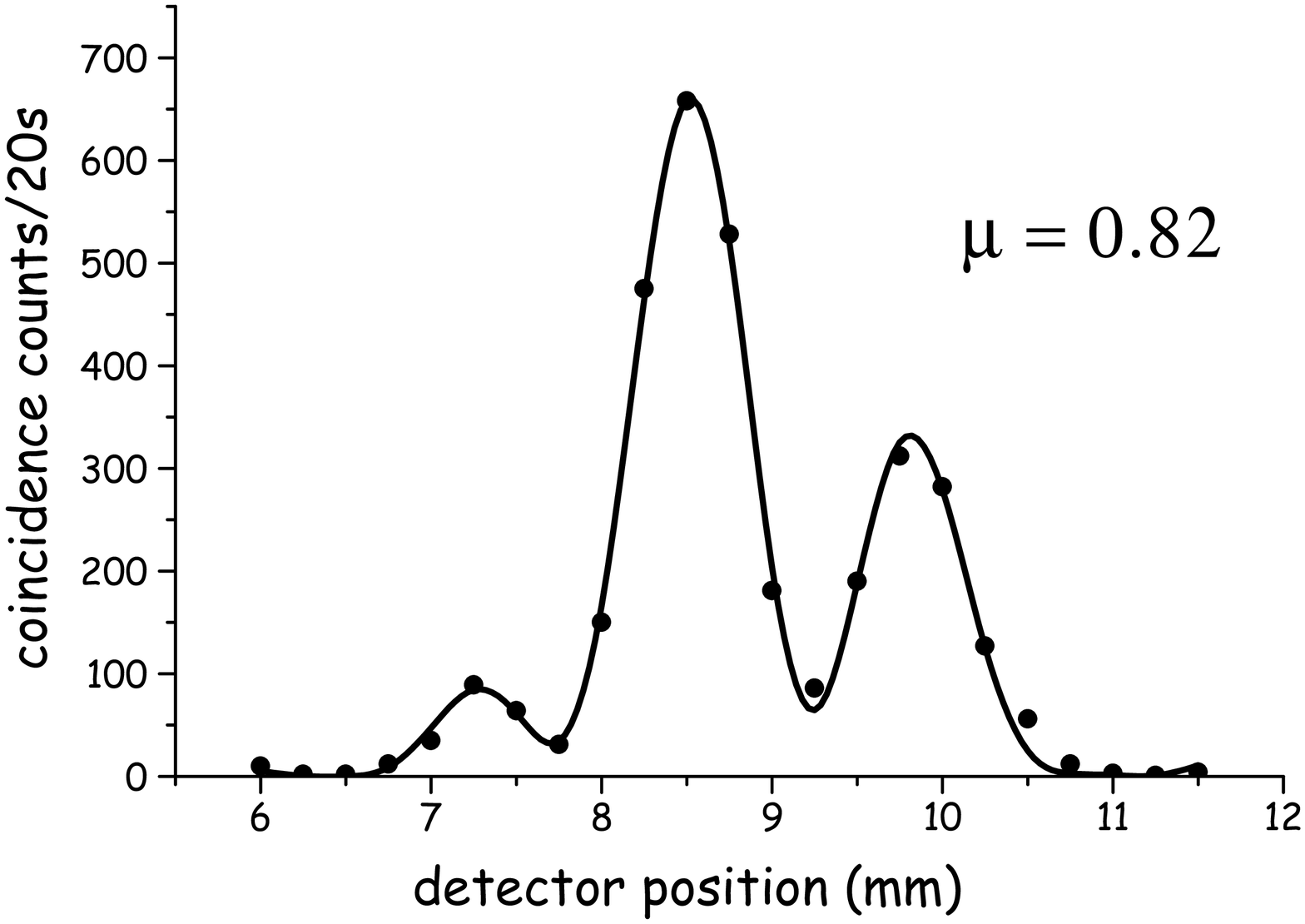,width=9cm,height=5cm}
\caption{Quantum interference for twin beams with different polarizations. High visibility
and high degree of polarization entanglement. Dots are experimental
data. The error bars are negligible. The solid line is a fitting to the usual double slit pattern function
giving the visibility.}
\label{fig4}
\end{figure}

\begin{figure}[h]
\vspace*{5cm}
\epsfig{file=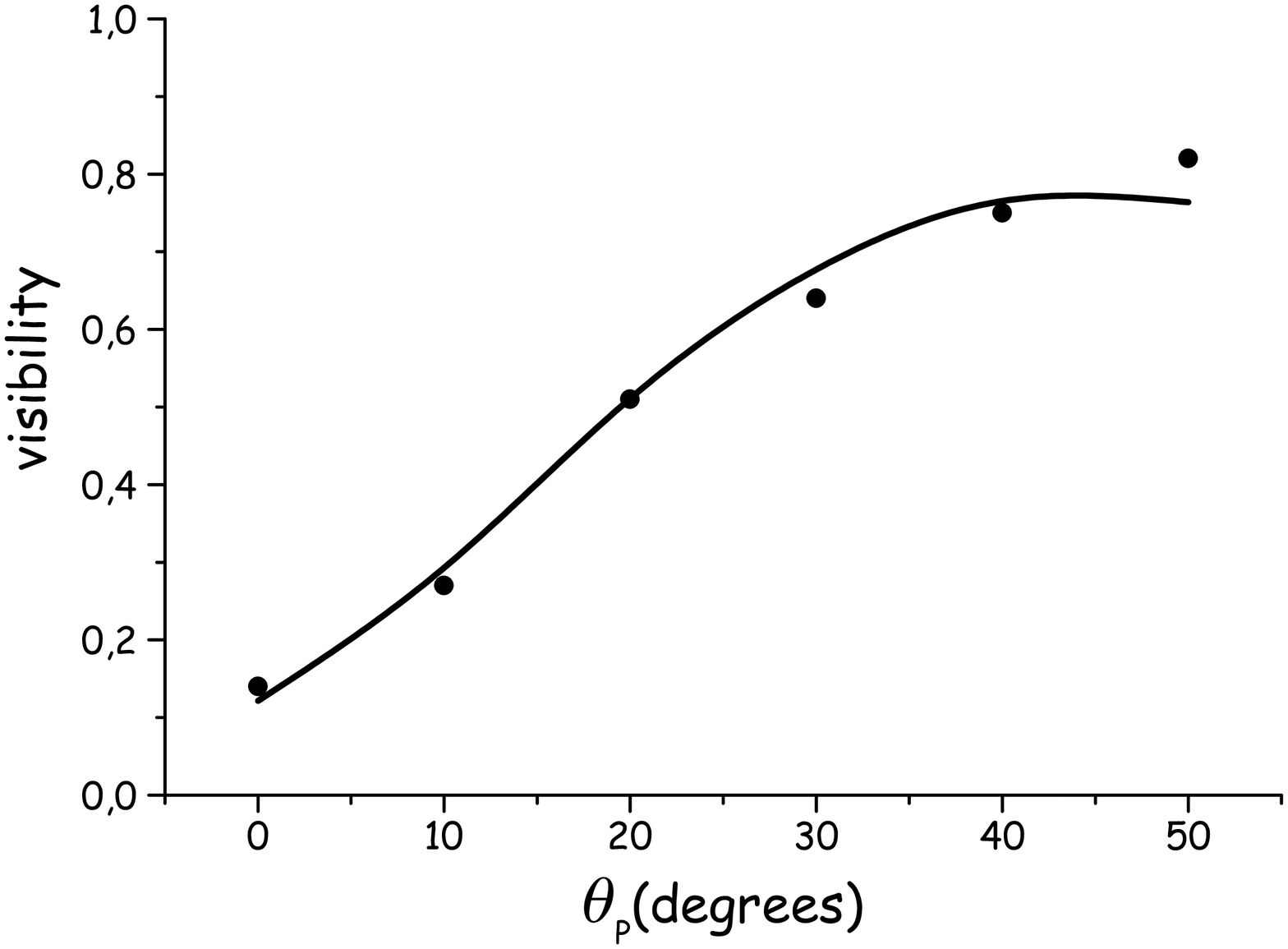,width=9cm,height=5cm}
\caption{Evolution of the visibility of the interference fringes when the pump beam 
polarization and the degree of entanglement is changed. Dots represent the visibilities
obtained from experimental data through nonlinear fitting. The solid line is a fitting to the
function described in Eq. (\ref{est8})}.
\label{fig5}
\end{figure}

We have carried out these measurements. The coincidence interference pattern for a non entangled state is shown in Fig.\ref{fig3} and the pattern for maximal entanglement is shown in Fig. \ref{fig4}. The lower visibility we have obtained, when only one crystal was actually pumped, was $\mu$ = 0.14
and the higher visibility, when both crystals were equally pumped, was $\mu$ = 0.82. In Fig. \ref{fig5} we present the evolution of the visibility of the interference fringes as a function of the pump beam polarization angle $\theta_{p}$.

\section{Discussion}

We showed that two parametric down-conversion crystals aligned in such a way that the generated twin-photon beams have good mode-matching, produce entangled states both in polarization and position, as in state (\ref{est1}). Although position is defined in the continuum, in our case only the macroscopic position of creation of each pair is relevant. Therefore, our system is defined in a 4x4 Hilbert space featuring two possible polarizations and origins for each pair. However, the emerging polarization is defined by the orientation of each nonlinear crystal. As a result, each pair presents a perfect correlation between both degrees of freedom, polarization and position(origin), reducing the effective Hilbert space to the usual 2x2. Depending on the relative orientation between the crystals, this state space can be solely the polarization Hilbert space, as in Eq. (\ref{est3}), when both crystals are oriented at orthogonal polarizations, or a double-slit kind of state, as in Eq. (\ref{est2}), when both crystals are oriented at the same direction. In each case, the measurement of the relevant degree of freedom will
reveal perfect quantum superposition. On the other hand, any measure on the complementary one will present no interference at all. 

In our experiment, we have taken advantage of the coupling between position and polarization
to switch from one kind of entanglement to another, according to our convenience. We have observed
that the interference due to the position entanglement is easy to measure. The resulting
interferometer is quite stable because the beams propagate together almost all the time and the
phase of the pattern can be easily varied by displacing the detectors. 
Thus, for example, once we have produced polarization entangled photons we could make them
entangled in position by the passage through 45$^{0}$ analyzers. Moreover, the measurement
of the position interference visibility has given us the degree of polarization entanglement of the
photon pair just after the crystals. The opposite could also have been done, for example, by preparing
position entangled photons and placing halfwave plates in the beams produced in crystal 1, but not
in those produced in crystal 2. Although more complicated from the experimental point of view,
this is another possibility of such a system.

The visibility of the interference fringes is limited by technical details. For example,
the detector entrance slit has a finite size and the fringes are smoothed, reducing the visibility.
The finite length of the crystals and their separation reduces the quality of the spatial mode
matching and consequently the visibility.
Anyway, the degree of entanglement in polarization can be obtained from those visibilities. In
our experiment, the maximal visibility observed was $\mu$ = 0.83 for the position interference
directly and $\mu$ = 0.82 for the polarization entangled state converted to position entangled 
state. The fact that they are about the same, shows the high fidelity in the conversion of 
polarization to position entanglement.

The lower visibility should be zero ideally. In our experiment it was $\mu$ = 0.14.
In this case, another technical detail has been responsible for the non-extinction of the
visibility. The polarization state of the pump beam is not exactly linear, but slightly elliptical.
Even if the eccentricity of the ellipsis is very small, the influence in the visibility is considerable
for small visibilities. For high visibilities however, the elliptical polarization has no influence.
The fitting of the curve for the evolution of the visibilities as a function of the pump beam
polarization (Fig.\ref{fig5}) has taken this problem into account. 
An elliptical polarization can be represented by the following state:
\begin{equation}
|\epsilon\rangle = [\epsilon_1 \cos\theta - i\epsilon_2 \sin\theta] |V\rangle + 
[\epsilon_1 \sin\theta + i\epsilon_2 \cos\theta] |H\rangle ,
\label{est6}
\end{equation} 
where $\theta = \theta_p$ is the angle between the vertical direction and the main
axis of the ellipsis.
Pumping photons in the above state, Eq. (\ref{est6}), lead to the following state for
the downconverted photons, in replacement of Eq. (\ref{est5}):

\begin{eqnarray}
| \Psi\rangle = [\epsilon_1 \cos\theta - i\epsilon_2 \sin\theta]|2,0\rangle + \\ \nonumber
e^{i\varphi }[\epsilon_1 \sin\theta + i\epsilon_2 \cos\theta]|0,2\rangle,
\label{est7} 
\end{eqnarray}
which gives rise to a coincidence interference pattern given by
\begin{equation}
C \propto \epsilon_{1}^2 + \epsilon_{2}^2 + (\epsilon_{1}^2 - \epsilon_{2}^2)\sin(2\theta)\cos\varphi
- 2\epsilon_1\epsilon_2 \sin\varphi.
\label{est8}
\end{equation}
Notice that, for example,  when $\epsilon_2 = 0$
we recover the coincidence profile obtained for linearly polarized pump C $\propto$ $\epsilon_{1}^2[1+\sin(2\theta)\cos\varphi]$. The effective visibility, which has been used to fit the experimental results, is given by
\begin{equation}
\mu_{eff}(\theta) = \mu_{max}\left[\frac{v_1^2}{v_2(\theta)}+v_2(\theta)\right]
\cos\left\{\mbox{arctg}\left[\frac{v_1}{v_2(\theta)}\right]\right\},
\label{est8} 
\end{equation}
where $v_1 = 4\epsilon_{1}^2(1-\epsilon_{1}^2)$ and $v_2(\theta)=(2\epsilon_{1}^2-1)\sin 2(\theta-\theta_0)$. $\mu_{max}$ is the higher visibility obtained experimentally. Setting it as a free
parameter we have found  $\mu_{max}$ = $0.77$ from our data. $\theta_0$ is a constant, which was also set as a free parameter and found to be nearly equal to $\pi$. $\epsilon_2$, which is
zero for a perfectly linearly polarized pump, was found to be 0.08. This shows that even small elliptical components in the pumping laser are enough to keep a residual visibility of the interference pattern.

It is easy to understand qualitatively
the role of the elliptical polarization on the
visibility of the interference patterns. The higher visibility occurs when the polarization of the
pump has projections onto the vertical and horizontal directions with the same magnitude.
This is achieved with a linear polarization at 45$^{0}$. This could also be accomplished
with a circularly polarized beam. Thus, as an elliptical state can be viewed as superposition of
a linear plus a circularly polarized state, we see that both components lead to the desired maximal entanglement.
The lower visibility is obtained when only one crystal is being pumped. This limit is reached only for a very pure linear polarization state. Any circularly polarized state component would lead
to a visibility higher than zero.

There are other means of measuring the degree of polarization entanglement of photon pairs. One of the most sophisticated (but also experimentally simple) is the quantum state tomography\cite{6}. It can also 
be measured by the visibility of the fringes obtained when one of the polarization analyzers is rotated and the other is kept fixed in 45$^0$, like in a Bell's inequality violation experiment.  
In every case it is possible to relate the degree of entanglement to a measurable parameter.
In our approach we stress the coupling between the two kinds of entanglement and their complementarity. 
We make use of this complementarity to measure one degree of freedom and extract information about the other.

\section{Conclusion}

In conclusion, we have presented a scheme capable of producing polarization and position entangled states for twin photon pairs. We have shown experimentally how each kind of entangled state can be generated from the same pair of crystals and how we can turn polarization entangled states into position entangled states. We have utilized this approach to measure the degree of polarization entanglement for twin photon pairs through position interference.

\section{ACKNOWLEDGMENT}
The authors  would like to acknowledge Prof. L. Davidovich for stimulating discussions and
the Brazilian agencies CNPq, PRONEX, FAPERJ and FUJB for the financial support.

\begin{thebibliography}{99}
\bibitem{11a}
This is the usual Young's experiment state, where state $|0,1\rangle$($|1,0\rangle$) was
replaced with $|0,2\rangle$($|2,0\rangle$).
\end{thebibliography}
\end{document}